\documentclass[twocolumn]{svjour3}          
\smartqed  
\usepackage{graphicx}
\usepackage{mathptmx}      
\usepackage{amsmath}
\usepackage{amssymb}
\usepackage{amsopn}
\usepackage{dsfont}
\usepackage{bm}
\usepackage{tikz}
\usepackage{tikz-3dplot}
\usepackage{epstopdf}
\usepackage{color}

\begin{document}

\title{Effects of noise-induced coherence on the performance of quantum absorption refrigerators}


\author{Viktor Holubec \and Tom{\'a}{\v s} Novotn{\'y}}


\institute{V. Holubec \at
 Department of Macromolecular Physics,  Faculty of Mathematics and Physics,  Charles University,
 V Hole{\v s}ovi{\v c}k{\' a}ch 2, 
 CZ-180~00~Prague, Czech Republic \\
 Tel.: +420-95155-2371\\
 Fax: +420-95155-2350\\
 \email{viktor.holubec@gmail.com} \\
 \emph{Present address:} 
 Institut f{\"u}r Theoretische Physik, 
 Universit{\"a}t Leipzig, 
 Postfach 100 920, D-04009 Leipzig, Germany
 \and
 T. Novotn{\'y} \at
 Department of Condensed Matter Physics, Faculty of Mathematics and Physics, Charles University, Ke Karlovu 5, CZ-121 16 Prague, Czech Republic\\
 Tel.: +420-95155-1392\\
 Fax: +420-22491-1061\\
 \email{tno@karlov.mff.cuni.cz} 
}

\date{Received: date / Accepted: date}

\maketitle

\begin{abstract}
We study two models of quantum absorption refrigerators with the main focus on discerning the role of noise-induced coherence on their thermodynamic performance. Analogously to the previous studies on quantum heat engines we find the increase in the cooling power due to the mechanism of noise-induced coherence. We formulate conditions imposed on the microscopic parameters of the models under which they can be equivalently described by classical stochastic processes and compare the performance of the two classes of fridges (effectively classical vs.~truly quantum). We find that the enhanced performance is observed already for the effectively classical systems, with no significant qualitative change in the quantum cases, which suggests that the noise-induced-coherence enhancement mechanism is caused by static interference phenomena.     

\end{abstract}


\section{Introduction}

The quantum coherence can be either used as a fuel of quantum heat engines \cite{Scully2003,Dillenschneider2009} or as an ingredient of non-equilibrium reservoirs which may yield efficiencies of thermodynamic machines exceeding the second law bounds \cite{Abah2014,Rosnagel2014,Klaers2017}. The common feature of all these approaches is that once the work needed to create the coherence is taken into account, the resulting thermodynamic efficiency becomes at best the one dictated by the second law.

In a recent series of papers, Scully et al.~\cite{Scully2011,Svidzinsky2011,Svidzinsky2012,Dorfman2013} reported a surprising enhancement of the thermodynamic performance (in particular the power output) of quantum heat engines with (nearly) degenerate levels, where the coherence is created `for free' just by connecting the system simultaneously to several different thermal environments. They named the mechanism the {\em noise-induced coherence}. This mechanism directed increased attention of the quantum thermodynamics community to quantum heat engines \cite{Tscherbul2014,Killoran2015,Xu2016,Dodin2016,Su2016,Streltsov2017} and raised many questions concerning its exact origin. Subsequent study \cite{Creatore2013} addressed the status of the used quantum optical master equation and found similar effects using just a Pauli rate equation, describing the working principle of Scully's model by a purely classical stochastic process (master equation where the coherences decouple from populations was derived after performing a polaron transformation). Recently, another study appeared treating a further refined model numerically, considering also non-Markovian effects due to the coupling of the system to a highly correlated environment \cite{Chen2016}. Still, the exact mechanism of the effect remains somewhat unclear which makes it difficult to predict if it will be relevant in other setups and what the essential ingredients of such setups might be. In particular, to the best of our knowledge, this mechanism has not been considered in the framework of quantum refrigerators.  

In this paper, we quite systematically address the presence and relevance of the noise-induced coherence in two model systems of quantum absorption refrigerators.~\footnote{In the quantum realm, the concept of absorption refrigerators, invented in 1858 by F. Carr{\'e}, experienced a renaissance initiated by Levy and Kosloff \cite{Levy2012} (see for example Refs.~\cite{Correa2013,Brask2015,Correa2014,Silva2015,Gonzalez2017,Hofer2018}). In all these studies, a specific model of quantum absorption refrigerator is investigated which cannot be mapped to the models considered by us. Inasmuch as we are aware, no model of absorption refrigerator which would contain degenerate energy levels and utilize the effect of noise-induced coherence has been studied in the literature yet.} In particular, we focus in detail on the following problem: Is the noise-induced coherence described by Scully and coworkers a true quantum effect, or can it be always equivalently described by a Pauli master equation? In the two models, we first study when the quantum optical master equation can be recast into a form in which coherences decouple from populations, i.e.~into the form of a Pauli master equation. We formulate microscopic conditions for which such a procedure is feasible. After that, using the Leggett-Garg inequalities \cite{Emary2014}, we put under a thorough test of quantumness the models which cannot be equivalently described as classical and, furthermore, we compare thermodynamic behavior and performance of the two classes of models --- those which are effectively classical and the true quantum ones.     

The outline of the paper is as follows. In Sec.~\ref{QME} we derive the optical master equation for systems with (nearly) degenerate energy levels with a special focus on conditions under which this master equation can be used. As far as we know, such a precise derivation is still missing in the literature leading to erroneous usage of the master equations in parameter regimes beyond the scope of their validity \cite{Creatore2013}. In Sec.~\ref{Definitions} we introduce the two model refrigerators and in Sec.~\ref{Quantum} and Appendix~\ref{Appx:LGI} we apply to them various criteria of quantumness. In Sec.~\ref{Thermodynamics} we introduce thermodynamic quantities and a thermodynamic description of the two models which we then study in detail both analytically and numerically in Sec.~\ref{Results}. Finally, we conclude our findings in Sec.~\ref{Conclusions}.


\section{Optical master equation for systems with (nearly) degenerate energy levels}\label{QME}

\subsection{Redfield master equation}
\label{sec:redfield}

\begin{figure}
\centerline{
    \begin{tikzpicture}[
      scale=0.5,
      level/.style={thick},
      virtual/.style={thick,densely dashed},
      trans/.style={thick,<->,shorten >=2pt,shorten <=2pt,>=stealth},
      classical/.style={thin,double,<->,shorten >=4pt,shorten <=4pt,>=stealth}
    ]  
		\draw[trans,red] (0cm,-3em) -- (-1.5cm,10.5em) node[midway,left] {\color{black} $\gamma_{1}$};
		\draw[trans,red] (0cm,-3em) -- (1.5cm,9.4em) node[midway,right] {\color{black} $\gamma_{2}$};
    \draw[level] (1.5cm,-3em) node[right,above] {$\left|3\right>$} -- (-1.5cm,-3em) node[left,below] {}; 
		\draw[level] (-3cm,+10.5em) node[right,below] {} -- (0cm,+10.5em) node[left,above] {$\left|1\right>$};
		\draw[level] (0cm,+9.4em) node[right,below] {} -- (3cm,+9.4em) node[left,above] {$\left|2\right>$};
    \end{tikzpicture}
}
\caption{V-type system used for deriving the quantum optical master equation (\ref{eq:3level_ini})-(\ref{eq:3level_end}). The upper doublet is not perfectly degenerate, $E_1-E_2=\Delta \hbar$. Only transitions depicted by the arrows are allowed.}
\label{fig:3level_system}
\end{figure}
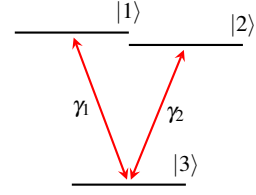

Consider the V-type three level system with nearly degenerate upper energy levels depicted in Fig.~\ref{fig:3level_system} in contact with a thermal environment in the form of a black body radiation field. The Hamiltonian of the system and the radiation,
\begin{equation}
\mathcal{H} = \mathcal{H}_S + \mathcal{H}_B + {\cal{H}}_{SB},
\label{eq:Full_Hamiltonian}
\end{equation}
is composed of three terms: the system Hamiltonian $\mathcal{H}_S = E_1\left|1\right>\left<1\right| + E_2\left|2\right>\left<2\right|$, the radiation bath Hamiltonian $\mathcal{H}_B = \hbar\sum_{\mathbf{k}} \nu_{\mathbf{k}} \hat{a}_{\mathbf{k}}^{\dagger}\hat{a}_{\mathbf{k}}$ and the coupling Hamiltonian ${\cal{H}}_{SB} = |e| \hat{\mathbf r}\cdot\hat{\mathbf E}$ describing the interaction of light and matter in the electric dipole approximation.

We assume that the coupling between the system and the reservoir is weak and that the relaxation time of the reservoir, $\tau_R$, is much shorter than the relaxation time of the system, $\tau_S$. Then the dynamics of the density matrix of the system alone ${\rho} = {\rho}(t)$, which can be calculated by tracing the full density matrix over the reservoir degrees of freedom, can be described up to the second order in the system-reservoir coupling strength by the so called Redfield master equation \cite{Cohen-Tannoudji1977}, \cite{Breuer2002}. In the interaction picture, this  reads
\begin{multline}
\dot{\tilde{\rho}}(t) = -\sum_i\sum_{\omega,\omega'} \exp\left({\rm i}(\omega-\omega')t\right)
\left[F(\omega) + {\rm i} S(\omega) \right]\times\\
\left[\mathbf{A}_i^\dagger(\omega')\mathbf{A}_i(\omega)\tilde{\rho}(t) - \mathbf{A}_i(\omega)\tilde{\rho}(t)\mathbf{A}_i^\dagger(\omega')\right] + {\rm c.c.}
\label{eq:MEQFUll}
\end{multline}
Here $\omega$ and $\omega'$ run over all allowed transition frequencies in the system (for the V-type model $\pm E_1/\hbar, \pm E_2/\hbar$), $S(\omega)$ denotes the Lamb and Stark shift terms and
\begin{equation}
F(\omega) =  \frac{|\omega|^3}{6\pi \hbar \epsilon_0 c^3} \left[(n(\omega) + 1)\Theta(\omega) + n(\omega)\Theta(-\omega) \right],
\label{eq:F}
\end{equation}
where the function
\begin{equation}
n(\omega) \equiv n(\omega, T)  = \frac{1}{\exp[\hbar \omega/k_B T] - 1}
\label{eq:n}
\end{equation}
gives the mean number of photons with frequency $\omega$ in the reservoir at temperature $T$. In Eqs.~(\ref{eq:F}) and (\ref{eq:n}), $\hbar$ stands for the Planck constant, $c$  the speed of light, $\epsilon_0$  the vacuum permittivity, and $k_B$  the Boltzmann constant. The symbol $\Theta(\omega)$ in Eq.~(\ref{eq:F}) denotes the unit-step (Heaviside) function. 

The vector operators $\mathbf{A}(\omega) = \mathbf{A}^{\dagger}(-\omega)$ in Eq.~(\ref{eq:MEQFUll}) are determined by allowed transitions in the system. For the V-type model of Fig.~\ref{fig:3level_system} they read $\mathbf{A}(E_1/\hbar) = \mathbf{g_{1}}\left|3\right>\left<1\right|$ and $\mathbf{A}(E_2/\hbar) = \mathbf{g_{2}}\left|3\right>\left<2\right|$. The generally complex-valued vectors $\mathbf{g_{1,2}}$ stand for elements $e\left<1\right|\hat{\mathbf r}\left|3\right>$ and $e\left<2\right| \hat{\mathbf r}\left|3\right>$ of the system electric dipole moment (with $e>0$ being the elementary charge), respectively.

The Redfield master equation is a Markovian master equation which is not of the Lindblad type and thus it in general does not preserve the positivity of the density matrix. Hence, before using this master equation, one usually brings it to the Lindblad form by further performing the so called rotating wave approximation (RWA) where all summands on the right hand side of Eq.~(\ref{eq:MEQFUll})  with $\omega \neq \omega'$ are neglected. The RWA approximation can be performed if the individual levels are well spaced such that the time-scales given by differences in the transition frequencies, $\tau_{RWA} = |\omega-\omega'|^{-1}$, are much shorter than the relaxation time of the system, $\tau_S$.  

Another frequent approximation lies in neglecting the Lamb and Stark shift terms $S(\omega)$. This approximation is reasonable since the Lamb and Stark shift terms are usually very small (for example for hydrogen the Lamb shift is of the order $10^{9}$ s$^{-1}$ \cite{Lamb1947} while typical optical transition frequencies are of the order of $10^{14}-10^{15}$ s$^{-1}$) and must be taken into account only in higher orders in the system-reservoir coupling strength. One example where the calculation is performed without dropping the Lamb and Stark shift terms is Ref.~\cite{BulnesCuetara2016}, where the authors investigate transport through a degenerate quantum dot. Different from Planck's spectral density $\omega^3/[\exp(\beta \hbar \omega) - 1]$ used here ($\beta\equiv1/k_{B}T$), the authors of the study \cite{BulnesCuetara2016} use the Ohmic spectral density $\omega/(\exp[\beta \hbar \omega]+1)$ which leads to weaker divergences in the Lamb and Stark shift terms. 

If the system contains close energy levels resulting in close transition frequencies $|\omega-\omega'|^{-1} = 1/\Delta \approx \tau_S$, the above approximations must be taken with care. First, the RWA approximation should be performed only in terms where $|\omega-\omega'|^{-1} \ll \tau_S$. This approximation is usually called the secular approximation \cite{BulnesCuetara2016} and it leaves on the right hand side of Eq.~(\ref{eq:MEQFUll}) the slowly rotating terms proportional to $\exp(\pm i \Delta t)$. In case of a non-perfect degeneracy ($\Delta \neq 0$), these terms cause several mathematical problems. First, the resulting master equation is still of the Redfield type and thus it does not preserve positivity of the system density matrix \cite{Creatore2013}. Second, the resulting master equation in the Schr{\" o}dinger picture contains terms proportional to $\Delta$. These terms can have similar magnitude as the Lamb and Stark shift terms and also the higher-order terms in the system-reservoir coupling. In order to account for a non-zero $\Delta$ correctly, one thus should not neglect the Lamb and Stark shift terms and should go beyond the week coupling Markov approximation of Eq.~(\ref{eq:MEQFUll}) at the same time. To sum up, Eq.~(\ref{eq:MEQFUll}) cannot be expected to give reasonable results for systems with close, but non-degenerate energy levels. In our analysis, we will thus always assume that the close energy levels are perfectly degenerate in which case there are no problems present.

\subsection{Summary of assumptions}
\label{sec:assumptions}

To sum up, for the derivation of the master equation for the V-type system with degenerate energy levels in the next subsection, we assume that the following separation of timescales occurs in the compound system composed of the system and reservoir:
\begin{eqnarray}
\tau_S &\gg& \tau_R,
\label{eq:condR}\\
\tau_S &\gg& \tau_{RWA}.
\label{eq:conRWA}
\end{eqnarray}
Here $\tau_S$ denotes the relaxation time of the system which is determined by the inverse maximum transition rate in the rate equation, $\tau_R$ denotes the relaxation time of the reservoir which is roughly given by $\tau_R = \hbar/(k_{B}T)$ \cite{Cohen-Tannoudji1977,Breuer2002,Carmichael2009}, and $\tau_{RWA}$ is determined by the smallest nonzero transition frequency in the system. The condition (\ref{eq:condR}) is necessary for the validity of the Markov approximation and the condition (\ref{eq:conRWA}) justifies usage of RWA approximation as described in Sec.~\ref{sec:redfield}. 

The inequalities (\ref{eq:condR}) and (\ref{eq:conRWA}) reduce the space of allowed parameters for the models based on the resulting master equation. The time scales $\tau_R$ and $\tau_{RWA}$ depend only on reservoir temperature and on the structure of energy levels of the system. Thus they can be determined a priori before the master equation is derived. On the other hand, the system relaxation time $\tau_S$ must be calculated a posteriori from the transition rates of the resulting master equation. The validity of inequalities (\ref{eq:condR}) and (\ref{eq:conRWA}) thus should be checked for every set of parameters separately\footnote{Let us note that Dorfman et al.~\cite{Dorfman2013} used the rotating wave approximation while assuming a $\Delta$ of the same order as the standard transition frequencies between the energy levels in their numerical examples (see Table 1 in Ref.~\cite{Dorfman2013}).}.

In addition to the described time scale separation, we assume that the system-bath coupling is weak (we use the second order approximation in the coupling strength), we neglect the Lamb and Stark shift terms and we assume that the closely-lying energy levels are actually degenerate. 

\subsection{Master equation for the V-type system with degenerate levels}\label{V-system}

Using these assumptions, the Redfield equation for the V-type model of Fig.~\ref{fig:3level_system} with degenerate upper doublet ($\Delta = 0$) results in the following 
set of coupled differential equations for the individual elements $\rho_{ij}\equiv \rho_{ij}(t)\equiv \left<i|\rho(t)|j\right>$ of the system density matrix in the Schr{\" o}dinger picture:
\begin{eqnarray}
\dot{\rho}_{11} &=& \gamma_1( f_{13}\rho_{33} - f_{31}\rho_{11}) - f_{32} {\rm Re}(\gamma_{12}^{c} \rho_{21}) ,
\label{eq:3levelF_ini}\\
\dot{\rho}_{22} &=& \gamma_2( f_{23}\rho_{33} - f_{32}\rho_{22}) - f_{31} {\rm Re}(\gamma_{12}^{c} \rho_{21}) ,\\
\dot{\rho}_{33} &=& \gamma_1(f_{31}\rho_{11} - f_{13}\rho_{33})
+ \gamma_2(f_{32}\rho_{22} - f_{23}\rho_{33}) \\
&+& (f_{31}+f_{32}){\rm Re}(\gamma_{12}^{c} \rho_{21}), 
\nonumber\\
\dot{\rho}_{12} &=& \frac{\gamma_{12}^{c}}{2}\left[
(f_{13}+f_{23})\rho_{33} - f_{32}\rho_{11} - f_{31}\rho_{22}\right]\\
&-& \frac{1}{2}(\gamma_1 f_{31}+\gamma_2 f_{32})\rho_{12}, 
\nonumber
\\
\dot{\rho}_{21} &=& \frac{{\gamma_{12}^{c}}^*}{2}\left[
(f_{13}+f_{23})\rho_{33} - f_{32}\rho_{11} - f_{31}\rho_{22}\right]
\nonumber
\\
&-& \frac{1}{2}(\gamma_1 f_{31}+\gamma_2 f_{32})\rho_{21}.
\label{eq:3levelF_end}
\end{eqnarray}
Here $\gamma_i \equiv |\mathbf{g_{i}}|^2$,  $\gamma_{12}^{c} = \mathbf{g}_1 \cdot \mathbf{g}_2^*$ , and $f_{ij} \equiv 2 F((E_i - E_j)/\hbar)$. The condition $E_1 = E_2$ implies $f_{13}=f_{23}$ and $f_{31}=f_{32}$. The structure of the equations guarantees the necessary condition $\rho_{12} = \rho_{21}^*$. The magnitude $\gamma_{12}$ of the coupling parameter $\gamma_{12}^{c}\equiv \gamma_{12}e^{i\phi}$ controls how strongly the coherence $\rho_{12}$ couples to populations. Using it we can get rid of the superfluous phase $\phi$ and simplify the above system into a more compact form 
\begin{eqnarray}
\dot{\rho}_{11} &=& \gamma_1( f_{13}\rho_{33} - f_{31}\rho_{11}) - \gamma_{12} f_{32} \rho_R ,
\label{eq:3level_ini}\\
\dot{\rho}_{22} &=& \gamma_2( f_{23}\rho_{33} - f_{32}\rho_{22}) - \gamma_{12} f_{31} \rho_R,\\
\dot{\rho}_{33} &=& \gamma_1(f_{31}\rho_{11} - f_{13}\rho_{33})
+ \gamma_2(f_{32}\rho_{22} - f_{23}\rho_{33}) 
 \nonumber\\
&+& \gamma_{12} (f_{31}+f_{32})\rho_R,
 \\
\dot{\rho}_R &=& \frac{\gamma_{12}}{2}\left[
(f_{13}+f_{23})\rho_{33} - f_{32}\rho_{11} - f_{31}\rho_{22}\right] 
\nonumber
\\
&-& \frac{1}{2}(\gamma_1 f_{31}+\gamma_2 f_{32})\rho_R,
\label{eq:3level_end}
\end{eqnarray}
with real $\rho_R \equiv \rho_{12}e^{-i\phi}=\rho_{21}e^{i\phi}$. The coupling $\gamma_{12}$ can attain any value from the interval $[0,\sqrt{\gamma_1 \gamma_2}]$ depending on the angle between the vectors $\mathbf{g_{1}}$ and $\mathbf{g_{2}}$. In accord with the nomenclature used in the studies \cite{Scully2011,Svidzinsky2011,Svidzinsky2012,Dorfman2013}, we will from now on call the case of the maximum value of the coupling parameter $\gamma_{12} = \sqrt{\gamma_1 \gamma_2}$ as the {\em limit of maximum coherence} and the limit of the zero value of the coupling parameter $\gamma_{12} = 0$ as the {\em classical} or {\em no-coherence limit}.

Equations (\ref{eq:3level_ini})--(\ref{eq:3level_end}) are sufficient to describe the dynamics of populations of the system and the energy currents throughout the system. The dynamics of the elements of the density matrix which are not contained in the above equations is trivial. 

Due to the linear nature of the Redfield equation (\ref{eq:MEQFUll}), using the formulas (\ref{eq:3level_ini})--(\ref{eq:3level_end}) one can write an analogous master equation for an arbitrary system containing pairs of degenerate levels where the individual transitions are coupled to an arbitrary number of independent reservoirs at different temperatures given that the assumptions described in Sec.~\ref{sec:assumptions} are fulfilled. It is enough to write the dynamical equations for all V-type, $\Lambda$-type and two-level subsystems (dynamical equations for a two-level system can be obtained from Eqs.~(\ref{eq:3level_ini})--(\ref{eq:3level_end}) if one forbids one of the transitions, i.e. after setting $\mathbf{g_1} = 0$ or $\mathbf{g_2} = 0$) composing the whole system in question and then sum up these sub-equations\footnote{The linearity of system (\ref{eq:3level_ini})--(\ref{eq:3level_end}) only holds if all $\gamma_{12}^{c}$'s of various bath couplings to a given energy doublet posses the same phase, which we assume here. If not, one has to resort back to the original full set of Eqs. (\ref{eq:3levelF_ini})--(\ref{eq:3levelF_end}).}. While the long-time solution of the resulting dynamical equation for a system coupled to a set of reservoirs at the same temperature is always given by the Boltzmann distribution where coherences vanish, the system attains a nontrivial steady state with nonzero coherences in case it is coupled at least to two reservoirs at different temperatures \cite{Scully2011,Svidzinsky2011,Svidzinsky2012,Dorfman2013,BulnesCuetara2016}. 

In the studies \cite{Scully2011,Svidzinsky2011,Svidzinsky2012,Dorfman2013} it was suggested that the coherence induced by the temperature gradient, the so called {\em noise-induced coherence}, can be utilized to enhance the output power of quantum heat engines. In what follows, we  
introduce two models of absorption refrigerators whose performance may be also enhanced by the noise-induced coherence. After these models are introduced, we first investigate whether the effect of the noise-induced coherence in these systems is a true quantum effect which can not be emulated by a classical stochastic system. Then, we discuss whether the cooling flux in the refrigerators can be enhanced by the noise-induced coherence similarly as the output power of heat engines in Refs.~\cite{Scully2011,Svidzinsky2011,Svidzinsky2012,Dorfman2013}.

\section{Refrigerators: Definitions of two models}\label{Definitions}


\begin{figure*}
\centerline{
    \begin{tikzpicture}[
      scale=0.5,
      level/.style={thick},
      virtual/.style={thick,densely dashed},
      trans/.style={thick,<->,shorten >=2pt,shorten <=2pt,>=stealth},
      classical/.style={thin,double,<->,shorten >=4pt,shorten <=4pt,>=stealth}
			]  
		\pgfmathsetmacro{\R}{0.7}
		\draw[fill=red,color=green] (-4cm,0cm) circle (\R); 
		\draw[->,thick, green] (-2cm,0cm) -- (-4cm + 2\R,0cm);
		\node[draw=none, thick, white] at (-4cm,0cm) {$T_m$};
		\node[draw=none, black] at (-2.7cm +\R,-0.5cm) {$-q_m$};
		\draw[fill=red,color=blue] (4cm,0cm) circle (\R); 
		\draw[->,thick, blue] (4cm - 2\R,0cm) -- (2cm,0cm);
		\node[draw=none, thick, white] at (4cm,0cm) {$T_c$};
		\node[draw=none, black] at (2.8cm -\R,-0.5cm) {$q_c$};
		\draw[fill=red,color=red] (0,4cm) circle (\R); 
		\node[draw=none, thick, white] at (0cm,4cm ) {$T_h$};
		\draw[->,thick, red] (0cm,4cm - 2\R) -- (0cm,2cm);
		\node[draw=none, black] at (0.5cm,2.7cm -\R) {$q_h$};
		\draw[fill=white,thick] (-2cm,-2cm) rectangle (2cm,2cm);
	  \draw[fill=black,thick] (-1cm,-1cm) rectangle (1cm,1cm);
		\node[draw=none, thick, white] at (0cm,0cm ) {fridge};
		\node[draw=none, black] at (0cm,-2.6cm ) {cavity/filter at $T=0$};
		\node[draw=none, black] at (-4cm,4cm ) {b)};
		\begin{scope}[shift={(13,0)}]
		\pgfmathsetmacro{\R}{0.7}
		\draw[fill=red,color=red] (-4cm,0cm) circle (\R); 
		\draw[->,thick, red] (-4cm + 2\R,0cm) -- (-2cm,0cm);
		\node[draw=none, thick, white] at (-4cm,0cm) {$T_h$};
		\node[draw=none, black] at (-2.7cm +\R,-0.5cm) {$q_h$};
		\draw[fill=red,color=blue] (4cm,0cm) circle (\R); 
		\draw[->,thick, blue] (4cm - 2\R,0cm) -- (2cm,0cm);
		\node[draw=none, thick, white] at (4cm,0cm) {$T_c$};
		\node[draw=none, black] at (2.8cm -\R,-0.5cm) {$q_c$};
		\draw[fill=green,thick] (-2cm,-2cm) rectangle (2cm,2cm);
	  \draw[fill=black,thick] (-1cm,-1cm) rectangle (1cm,1cm);
		\node[draw=none, thick, white] at (0cm,0cm ) {fridge};
		\node[draw=none, black] at (0cm,-2.6cm ) {cavity/filter at $T=T_m$};
		\node[draw=none, black] at (0.8cm,-1.5cm) {$-q_m$};
		\draw[->,thick, black] (0cm,-1cm) -- (0cm,-1.5cm);
		\draw[->,thick, black] (-1cm,-1cm) -- (-1.5cm,-1.5cm);
		\draw[->,thick, black] (-1cm,0cm) -- (-1.5cm,0cm);
		\draw[->,thick, black] (-1cm,1cm) -- (-1.5cm,1.5cm);
		\draw[->,thick, black] (1cm,1cm) -- (1.5cm,1.5cm);
		\draw[->,thick, black] (0cm,1cm) -- (0cm,1.5cm);
		\draw[->,thick, black] (1cm,0cm) -- (1.5cm,0cm);
		\draw[->,thick, black] (1cm,-1cm) -- (1.5cm,-1.5cm);
		\node[draw=none, black] at (-4cm,4cm ) {c)};
		\end{scope}
		\begin{scope}[shift={(5,7)}]
		\draw[trans,green] (0cm,-3em) -- (-2.0cm,19.5em) node[midway,left] {\color{black} $\gamma_{1m}$};
		\draw[trans,green] (0cm,-3em) -- (2.0cm,19.5em) node[midway,right] {\color{black} $\gamma_{2m}$};
    \draw[trans,blue] (1.5cm,-3em) -- (5.9cm,2.7em) node[midway,right] {\color{black} $\gamma_c$};
		\draw[trans,red] (2.0cm,19.5em) -- (5.9cm,2.7em) node[midway,right] {\color{black} $\gamma_{2h}$};
		\draw[trans,red] (-2.0cm,19.5em) -- (5.9cm,2.7em) node[near end,left] {\color{black} $\gamma_{1h}$};
    \draw[level] (1.5cm,-3em) node[right,above] {$\left|3\right>$} -- (-1.5cm,-3em) node[left,below] {};
		\draw[level] (-3.5cm,+19.5em) node[right,below] {} -- (-0.5cm,+19.5em) node[left,above] {$\left|1\right>$};
		\draw[level] (0.5cm,+19.5em) node[right,below] {} -- (3.5cm,+19.5em) node[left,above] {$\left|2\right>$};
		\draw[level] (7.4cm,+2.7em) node[right,above] {$\left|4\right>$} -- (4.4cm,+2.7em) node[left,above] {};
		\node[draw=none, black] at (-5cm,5cm ) {a)};
    \end{scope}
    \end{tikzpicture}
}
\caption{a) Scheme of the level structure of the working medium of the two refrigerators depicted in panels b) (A-type) and c) (B-type).}
\label{fig:fridges}
\end{figure*}
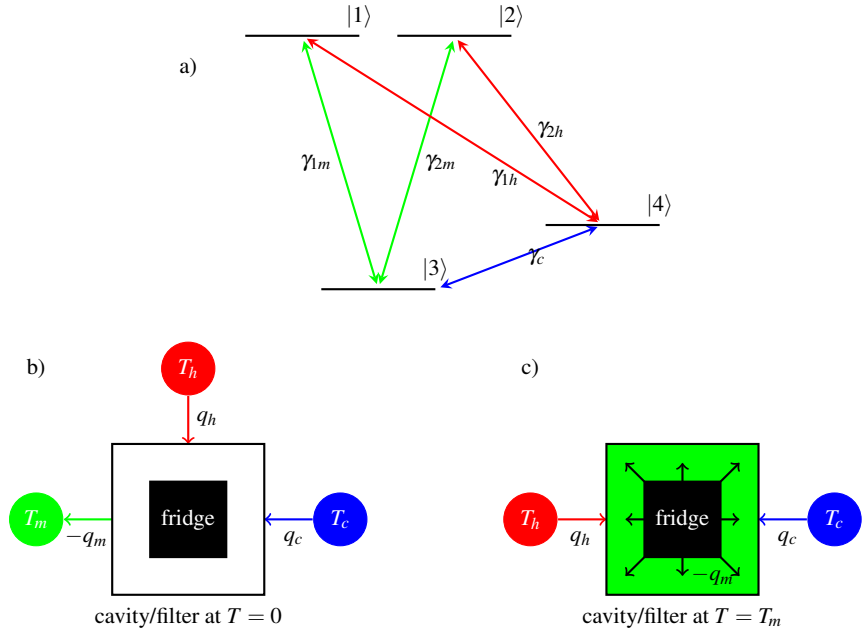

Consider the two types of absorption refrigerators depicted by general thermodynamic diagrams in Figs.~\ref{fig:fridges}b and \ref{fig:fridges}c. For both types of the refrigerators we assume the structure of energy levels of the working medium (system) depicted in Fig.~\ref{fig:fridges}a. Both refrigerators are coupled to three reservoirs at the temperatures $T_c<T_m<T_h$ (cold-medium-hot) and use the temperature gradient $T_h - T_m$ to cool the coldest bath. The reservoir at $T_h$ thus serves as an energy source of the fridge and the reservoir at $T_m$ is an entropy sink which collects the heat coming both from the hot and cold baths (see the arrows in Figs.~\ref{fig:fridges}b and \ref{fig:fridges}c).

The allowed transitions between the individual levels of the system are in Fig.~\ref{fig:fridges}a depicted by the colored arrows. For the A-type refrigerator, we assume that the transitions depicted by the individual colors are caused by the photons coming from heat reservoirs depicted by the same color in Fig.~\ref{fig:fridges}b. Naturally, if one would just couple the system to all three reservoirs, each of them would be in principle able to trigger all possible transitions. The selective coupling of the individual reservoirs to the specific transitions can be achieved only by using a suitable spectral filter which would allow the individual colored baths in Fig.~\ref{fig:fridges}b to interact solely with the correspondingly colored transitions in Fig.~\ref{fig:fridges}a. In practice, such a filter would have a non-zero temperature and thus it would also trigger  transitions in the system. For the A-type refrigerator we assume that the filter temperature is vanishingly small and thus it cannot cause any transition. At a first glance, such a refrigerator cheats a bit, because in fact it contains an additional reservoir at a lower temperature than that of the cold bath. In such a case, one actually does not need the fridge to cool the cold bath, it would be enough to connect it directly to the low-temperature filter.

Although the A-type fridge model suffers from the above-discussed issue with the coupling to the reservoirs, such a selective coupling is standardly assumed in the literature (see for example \cite{Scully2011,Svidzinsky2011,Svidzinsky2012,Dorfman2013,Joulain2016}). However, in those studies it is assumed that the system is composed of several interacting subsystems which are coupled to the individual reservoirs. In this case, one does not need to consider a filter in order to reason for the selective coupling. In the following, we will thus assume that the working medium (system) of the A-type fridge is of this kind, i.e. that it is composed of several subsystems each coupled to the individual bath.

To introduce also a more realistic fridge with a working medium consisting of a single system, we assume that the filter in the B-type fridge of Fig.~\ref{fig:fridges}c actually represents the reservoir at the temperature $T_m$. This means that all the transitions in the system can be induced by this bath. In addition to this, the red and blue transitions in Fig.~\ref{fig:fridges}a can be triggered by the hot bath ($T_h$) and by the cold bath ($T_c$), respectively.


In Fig.~\ref{fig:fridges}a, the individual parameters $\gamma_{1m}$, $\gamma_{2m}$, $\gamma_{1h}$, $\gamma_{2h}$ and $\gamma_c$ denote squared magnitudes  of the electric dipole moment elements corresponding to the given transitions. The system is composed of two V-type systems $\left|3\right>$-$\left|1\right>$-$\left|2\right>$ and $\left|4\right>$-$\left|1\right>$-$\left|2\right>$ and one two-level system $\left|3\right>$-$\left|4\right>$. We thus have two parameters measuring the coupling of the coherence $\rho_{12}$ to the populations. The first one, $\gamma_{12m} \in [0, \sqrt{\gamma_{1m} \gamma_{2m}}]$, stems from the V-type system $\left|3\right>$-$\left|1\right>$-$\left|2\right>$ and the second one,  $\gamma_{12h} \in [0, \sqrt{\gamma_{1h} \gamma_{2h}}]$, corresponds to the second V-type system $\left|4\right>$-$\left|1\right>$-$\left|2\right>$. For the A-type refrigerator, the mean number of photons $n(\omega)$ corresponding to the individual transitions can be obtained just by using the temperatures of the corresponding reservoirs in Eq.~(\ref{eq:n}). For the B-type refrigerator, the mean number of photons for a given transition is given by the sum over $n(\omega)$ corresponding to the individual reservoirs causing the transition. For example, for the transition $\left|4\right>$-$\left|1\right>$ we have $n_{\rm eff}(\omega_{14}) = n(\omega_{14}, T_h) + n(\omega_{14}, T_m)$, where $\omega_{14} \equiv (E_1-E_4)/\hbar$.

Using these parameters and  formulas (\ref{eq:3level_ini})--(\ref{eq:3level_end}) one can write rate equations for the constituent 2- and 3-level subsystems of the full refrigerator model. For the sake of simplicity, we consider only the situation where the phases $\phi$ corresponding to the two 3-level subsystems equal and thus one can still use a single $\rho_R \equiv \rho_{12}e^{-i\phi}=\rho_{21}e^{i\phi}$. Summing up the resulting equations, one obtains the total rate equation for the whole refrigerator. The resulting master equation for the system depicted in Fig.~\ref{fig:fridges}a reads:

\begin{eqnarray}
\dot{\rho}_{11} &=&   
   \tilde{\gamma}_{1m} n_m \rho_{33} + \tilde{\gamma}_{1h} n_h \rho_{44}-[\tilde{\gamma}_{1h} (n_h + 1) + \tilde{\gamma}_{1m} (n_m + 1)]\rho_{11}- \nonumber\\ 
	&-& [\tilde{\gamma}_{12h} (n_h + 1) + \tilde{\gamma}_{12m} (n_m + 1)]\rho_{R},
\label{eq:freq1}\\
\dot{\rho}_{22} &=&
   \tilde{\gamma}_{2m} n_m \rho_{33} + \tilde{\gamma}_{2h} n_h \rho_{44} - [\tilde{\gamma}_{2h} (n_h + 1) +\tilde{\gamma}_{2m} (n_m + 1)]\rho_{22} - \nonumber \\
	&-& [\tilde{\gamma}_{12h} (n_h + 1) + \tilde{\gamma}_{12m} (n_m + 1)] \rho_{R},
\label{eq:freq2}\\
\dot{\rho}_{33} &=& \tilde{\gamma}_{c} (n_c+1) \rho_{44} + \tilde{\gamma}_{1m} (n_m + 1) \rho_{11} + \tilde{\gamma}_{2m} (n_m + 1) \rho_{22} - \nonumber \\ &-&  (\tilde{\gamma}_{1m} n_m  + \tilde{\gamma}_{2m} n_m + \tilde{\gamma}_c n_c) \rho_{33} +  
 2\tilde{\gamma}_{12m} (n_m + 1) \rho_{R},
\label{eq:freq3}\\
\dot{\rho}_{44} &=& \tilde{\gamma}_c n_c \rho_{33} + \tilde{\gamma}_{1h} (n_h + 1) \rho_{11} + \tilde{\gamma}_{2h} (n_h + 1) \rho_{22}  - \nonumber \\
	&-&	[\tilde{\gamma}_{1h} n_h + \tilde{\gamma}_{2h} n_h + \tilde{\gamma}_c (1 + n_c)] \rho_{44} + 2\tilde{\gamma}_{12h} (n_h + 1) \rho_{R} ,
\label{eq:freq4}\\
\dot{\rho}_R &=& \frac{\tilde{\gamma}_{12m}}{2}[
2 n_m \rho_{33} - (n_m + 1) (\rho_{11} + \rho_{22})
]
\nonumber
\\
&+&
\frac{\tilde{\gamma}_{12h}}{2}[
2 n_h \rho_{44} - (n_h + 1) (\rho_{11} + \rho_{22})
]
- \nonumber \\ &-&
\frac{1}{2} [(\tilde{\gamma}_{1h} + \tilde{\gamma}_{2h}) (n_h + 1) + (\tilde{\gamma}_{1m} + \tilde{\gamma}_{2m}) (n_m + 1)]\rho_{R}.
\label{eq:freq6}
\end{eqnarray}
Different from Eqs.~(\ref{eq:3level_ini})--(\ref{eq:3level_end}) we now write the transition rates using the average number of photons $n_i$ corresponding to the individual transitions and the redefined parameters $\tilde \gamma_i = \tilde \gamma(\omega) = |\omega|^3\gamma_i/(6 \pi \hbar \epsilon_0 c^3)$, where $\omega$ denotes the frequency of the corresponding transition. For example $\tilde \gamma_{1m} = (\omega_1-\omega_3)^3 \gamma_{1m} /(6 \pi \hbar \epsilon_0 c^3)$ and similarly for other parameters $\tilde \gamma$. 
For the A-type fridge, we obtain the average photon numbers $n_h = n(\omega_{14},T_h)$, $n_m = n(\omega_{13},T_m)$ and $n_c = n(\omega_{43},T_c)$, where $\omega_{ij} \equiv (E_i-E_{j})/\hbar$. For the B-type fridge, the coefficient $n_m$ is the same as for the A-type fridge and the coefficients $n_h$ and $n_c$ differ due to the omnipresent action of the reservoir at $T_m$, $n_h = n(\omega_{14},T_m) + n(\omega_{14},T_h)$ and $n_c = n(\omega_{43},T_c) + n(\omega_{43},T_m)$. 

To close this section, we note that in order to calculate the steady-state density matrix $\rho$, it is advantageous to rewrite the system (\ref{eq:freq1})--(\ref{eq:freq6}) in the matrix form
\begin{equation}
\dot{\mathbf{\rho}} = \mathcal{L} \mathbf{\rho},
\label{eq:Mrate_EQ}
\end{equation}
where the matrix $\mathcal{L}$ contains the individual transition rates from Eqs.~(\ref{eq:freq1})--(\ref{eq:freq6}) and the vector $\mathbf{\rho} = \mathbf{\rho}(t)$ contains the matrix elements in question, $\mathbf{\rho} = (\rho_{11}, \rho_{22}, \rho_{33}, \rho_{44}, \rho_R)^{T}$. Having described the master equation for the two fridges, let us now discuss to what extent this master equation describes a pure quantum process.

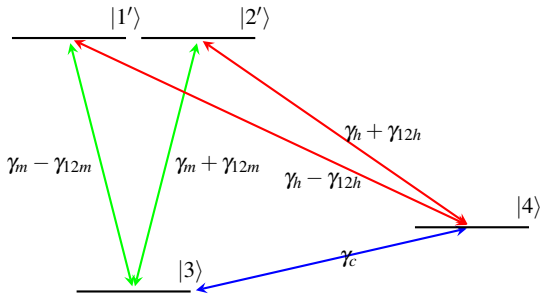
\begin{figure}
\centerline{
    \begin{tikzpicture}[
      scale=0.5,
      level/.style={thick},
      virtual/.style={thick,densely dashed},
      trans/.style={thick,<->,shorten >=2pt,shorten <=2pt,>=stealth},
      classical/.style={thin,double,<->,shorten >=4pt,shorten <=4pt,>=stealth}
    ]  
		\draw[trans,green] (0cm,-3em) -- (-1.7cm,19.5em) node[midway,left] {\color{black} $\gamma_{m}-\gamma_{12m}$};
		\draw[trans,green] (0cm,-3em) -- (1.7cm,19.5em) node[midway,right] {\color{black} $\gamma_{m}+\gamma_{12m}$};
    \draw[trans,blue] (1.5cm,-3em) -- (8.9cm,2.7em) node[midway,right] {\color{black} $\gamma_c$};
		\draw[trans,red] (1.7cm,19.5em) -- (8.9cm,2.7em) node[midway,right] {\color{black} $\gamma_{h}+\gamma_{12h}$};
		\draw[trans,red] (-1.7cm,19.5em) -- (8.9cm,2.7em) node[near end,left] {\color{black} $\gamma_{h}-\gamma_{12h}$};
    \draw[level] (1.5cm,-3em) node[right,above] {$\left|3\right>$} -- (-1.5cm,-3em) node[left,below] {}; 
		\draw[level] (-3.2cm,+19.5em) node[right,below] {} -- (-0.2cm,+19.5em) node[left,above] {$\left|1'\right>$};
		\draw[level] (0.2cm,+19.5em) node[right,below] {} -- (3.2cm,+19.5em) node[left,above] {$\left|2'\right>$};
		\draw[level] (10.4cm,+2.7em) node[right,above] {$\left|4\right>$} -- (7.4cm,+2.7em) node[left,above] {};
    \end{tikzpicture}
}
\caption{Scheme of the transition rates in the working medium of the two refrigerators depicted in Fig.~\ref{fig:fridges} in the transformed basis where coherences decouple from populations.}
\label{fig:system_fridge}
\end{figure}

\section{To be or not to be quantum: Change of the basis}\label{Quantum}

An important implication of the degeneracy of the levels $\left| 1\right>$ and $\left| 2\right>$ is that there is no a priory choice of the basis in the subspace $\left| 1\right>$, $\left| 2\right>$. 
The main aim of this subsection is to investigate whether one can find a basis in which the ``quantum" terms coupling the populations $\rho_{ii}$ to the coherence $\rho_R$ in the master equation (\ref{eq:freq1})--(\ref{eq:freq6}) would vanish. 

These terms are given by $\gamma_{12m} = |\mathbf{g_{1m}} \cdot \mathbf{g^{*}_{2m}}|$
and $\gamma_{12h} = |\mathbf{g_{1h}} \cdot \mathbf{g^{*}_{2h}}|$, where $\mathbf{g_{1m}} = e\left<1\right|\hat{\mathbf{r}}\left|3\right>$, $\mathbf{g_{2m}} = e \left<2\right|\hat{\mathbf{r}}\left|3\right>$, $\mathbf{g_{1h}} = e \left<1\right|\hat{\mathbf{r}}\left|4\right>$, and $\mathbf{g_{2h}} = e \left<2\right|\hat{\mathbf{r}}\left|4\right>$. At this point, let us assume that the scalar products of the matrix elements of the dipole moment operator are real which allows us to consider only real coordinate transformations\footnote{A common nonzero phase of the two scalar products, which we have assumed in Sec.~\ref{V-system}, can be absorbed into a complex transformation of the basis.}. To sum up, we  look for a new orthonormal basis $\left|1'\right> = -\sin\theta\left|1\right> + \cos\theta\left|2\right>$, $\left|2'\right> = \cos\theta \left|1\right> + \sin\theta \left|2\right>$ which would give $\gamma_{1'2'm} = 0$ and, simultaneously, $\gamma_{1'2'h} = 0$ for $\gamma_{12m} \neq 0$ and $\gamma_{12h} \neq 0$. We find that in the new basis the constants $\gamma_{1h,m}$, $\gamma_{2h,m}$ and $\gamma_{12 h,m}$ read
\begin{eqnarray}
\gamma_{1'h,m} &=& \gamma_{1h,m} \sin^2 \theta+ \gamma_{2h,m} \cos^2 \theta-  \gamma_{12 h,m}\sin 2\theta\label{eq:transQ1}\\
\gamma_{2'h,m} &=&  \gamma_{1h,m} \cos^2 \theta+ \gamma_{2h,m} \sin^2 \theta+  \gamma_{12 h,m}\sin 2\theta,
\label{eq:transQ2}\\
\gamma_{1'2' h,m} &=& \left|\frac{ \gamma_{2h,m}-\gamma_{1h,m}}{2} \sin2\theta+ \gamma_{12h,m}\cos 2\theta\right| .
\label{eq:transQ3}
\end{eqnarray}
The transformation from the original basis $\{\left|1\right>, \left|2\right>\}$ to the new one $\{\left|1'\right>, \left|2'\right>\}$
manifests itself in Eqs.~(\ref{eq:freq1})--(\ref{eq:freq6}) as a simple replacement of the constants  $\gamma_{1h,m}$, $\gamma_{2h,m}$, and $\gamma_{12 h,m}$ by $\gamma_{1'h,m}$, $\gamma_{2'h,m}$, and $\gamma_{1'2' h,m}$. The transformed master equation would not contain quantum terms, i.e.~coherences would decouple from occupations, if the right-hand side of Eqs.~(\ref{eq:transQ3}) vanished for both $\gamma_{1'2' h}$ and $\gamma_{1'2' m}$. For real transformations and real $\gamma_{12h,m}$ as considered here, this leads to two equations for one unknown variable $\theta$ (for complex transformations and complex $\gamma_{12h,m}$ this would lead to four equations for two unknown parameters of the transformation). 

If we rule out the unlikely case that $(\gamma_{1m}-\gamma_{2m})/\gamma_{12m}  = (\gamma_{1h}-\gamma_{2h})/\gamma_{12h} \neq 0$, we find out that the two equations can be simultaneously solved with the same $\theta$ only if $\gamma_{1h} = \gamma_{2h}( \equiv \gamma_{h})$ {\em and} $\gamma_{1m} = \gamma_{2m}( \equiv \gamma_{m})$. The corresponding value of $\theta$ is $\pi/4$ and the transformed coefficients (\ref{eq:transQ1})--(\ref{eq:transQ3}) read $\gamma_{1'h,m} = \gamma_{h,m} - \gamma_{12h,m}$, $\gamma_{2'h,m} = \gamma_{h,m} + \gamma_{12h,m}$, and $\gamma_{1'2' h,m} = 0$ as required. The system in the transformed basis is depicted in Fig.~\ref{fig:system_fridge}. It turns out that the coherence suppresses transitions to one of the transformed upper levels (subradiant; prefactors of the rate $\gamma_{h,m} - \gamma_{12h,m}$) and speeds up transitions to the other upper level (superradiant; prefactors of the rate $\gamma_{h,m} + \gamma_{12h,m}$). 

In the transformed basis, the limit of maximum coherence $\gamma_{12h,m} = \gamma_{h,m}$ thus leads to complete decoupling of one of the upper levels from the rest of the system (formation of a dark state). As a result, there are two stationary solutions of the transformed system (\ref{eq:freq1})--(\ref{eq:freq6}). The first one, $\mathbf{\rho} = (1,0,0,0,0)$, corresponds to state $\left|1'\right>$ which is fully decoupled from the reservoirs and consequently is called the {\em dark state} (as it cannot emit or absorb photons). The second stationary solution represents a standard non-equilibrium steady state in the subsystem formed by the levels $\left|2'\right>$,$\left|3\right>$ and $\left|4\right>$ coupled to the three reservoirs. The decoupling of one of the upper levels from the rest of the system is a manifestation of a solely quantum effect of the Fano interference. It is interesting that the systems exhibiting this purely quantum effect ($\gamma_{1h,m} = \gamma_{2h,m} = \gamma_{h,m}$ and $\gamma_{12h} = \gamma_{h,m}$) and the corresponding classical system depicted in Fig.~\ref{fig:system_fridge} exhibit the same dynamics (and thus also thermodynamics).

In cases $\gamma_{1h,m} \neq \gamma_{2h,m}$ when the two coefficients $\gamma_{1'2' h,m}$ cannot be simultaneously nullified, one can still find a coordinate transformation in which the coherence vanishes in the steady state and thus the stationary density matrix is diagonal. However, in such a case the populations and coherences are still coupled in the master equation and thus the dynamics of these systems can not be described by a classical stochastic process. Nevertheless, the finding of Fig.~\ref{fig:system_fridge} that one of the transition channels is always enhanced by the coherences and the other one is suppressed is still valid. Only the suppression/enhancement depends in a more complicated way on the individual parameters of the model.

To sum up, we have found that the dynamics of the system depicted in Fig.~\ref{fig:fridges}a can be described by a classical stochastic process in case of equal magnitudes of the dipole moment elements corresponding to the individual transitions, $\gamma_{1h,m} = \gamma_{2h,m}$. For different magnitudes $\gamma_{1h,m} \neq \gamma_{2h,m}$ and nonzero coefficients $\gamma_{12 h,m}$ the populations and coherences are always connected from the dynamical point of view. In Appendix~\ref{Appx:LGI}, we checked that a system exhibiting the noise-induced coherence in the latter parameter regime does not break the stationary Legget-Garg
inequalities \cite{Emary2014,Friedenberger2017} and thus it is not obvious whether it can be mimicked by a classical stochastic process, or not.

Having discussed possible issues concerning quantumness of the noise-induced coherence, let us now turn to the discussion of the thermodynamic properties of the refrigerators with degenerate energy levels.

\section{Thermodynamics}\label{Thermodynamics}


Our aim in this section is to determine heat flows from the individual reservoirs coupled to a system containing V-type transitions depicted in Fig.~\ref{fig:3level_system}. To do this, we must first identify probability currents between the individual energy levels of this system. In the system of Fig.~\ref{fig:3level_system}, each of the upper levels is connected solely to the site $\left|3\right>$ and thus the time derivatives of the upper-level populations are determined by the probability currents $j_{13}$ and $j_{23}$ from the site $\left|3\right>$ to the individual sites $\left|1\right>$  and $\left|2\right>$ : $\dot{\rho}_{11} = j_{13} = - j_{31} $ and $\dot{\rho}_{22} = j_{23} = - j_{32}$, where (see Eqs.~(\ref{eq:3level_ini})--(\ref{eq:3level_end}))
\begin{eqnarray}
j_{13} &=& \gamma_1( f_{13}\rho_{33} - f_{31}\rho_{11}) - f_{32} \gamma_{12} \rho_R,
\label{eq:j31}\\
j_{23} &=& \gamma_2( f_{23}\rho_{33} - f_{32}\rho_{22}) - f_{31} \gamma_{12} \rho_R.
\label{eq:j32}
\end{eqnarray}

These probability currents coincide with the  results of the analysis used in Ref.~\cite{BulnesCuetara2016}. Comparing them with the classical formulas following from the Pauli equation, one may conclude that the effect of coherences is contained solely in the terms proportional to $\gamma_{12}$. Nevertheless, in Eqs.~(\ref{eq:3level_ini})--(\ref{eq:3level_end}) the non-diagonal elements of the density matrix influence also populations and thus such a simple partition of the current into a classical and a non-classical part is not possible. A way to compare the results of the models with and without noise-induced coherence is thus to compare solutions for nonzero $\gamma_{12}$ (dynamics with coherences) and $\gamma_{12} = 0$ (classical dynamics). 

Using the formulas (\ref{eq:j31}) and (\ref{eq:j32}), one can write probability currents between arbitrary two-levels of the composite system of Fig.~\ref{fig:fridges}a. These currents can be also written comparing the master equation (\ref{eq:freq1})--(\ref{eq:freq6}) with its current conserving counterpart
\begin{eqnarray}
\dot{\rho}_{11} &=&  - j_{31} + j_{14},
\label{eq:freqJ1}\\
\dot{\rho}_{22} &=& - j_{32} + j_{24},
\label{eq:freqJ2}\\
\dot{\rho}_{33} &=& j_{31} + j_{32} - j_{43},
\label{eq:freqJ3}\\
\dot{\rho}_{44} &=& - j_{14} - j_{24} + j_{43}.
\label{eq:freqJ5}
\end{eqnarray}
In the steady state, the condition that the system density matrix is constant in time ($\dot{\rho}_{ij} = 0$) together with the probability conservation implied by the structure of the master equation (\ref{eq:freqJ1})--(\ref{eq:freqJ5}) give the continuity equation for the probability current throughout the system:
\begin{equation}
j_{43} = j_{14} + j_{24} =  j_{31} + j_{32} = j.
\label{eq:continuity}
\end{equation}
The positive direction of the probability current $j$ is chosen counter-clockwise, i.e.~ from 3 to 4.

In case the system is coupled to more than one heat bath, one can identify probability currents corresponding to the heat flows form the individual reservoirs. For example consider the transition from level $\left|4\right>$ to level $\left|1\right>$ which is in the B-type fridge caused simultaneously by the bath at temperature $T_m$ and the bath at temperature $T_h$, $n_h = n(\omega_{14},T_m) + n(\omega_{14},T_h)$. The part of the probability current $j_{14}$ caused by the bath at $T_x$ is then given by
\begin{multline}
j_{14;x} = \tilde{\gamma}_{1h}\{n(\omega_{14},T_x)\rho_{44} -[n(\omega_{14},T_x) + 1]\rho_{11}\}\\
-\tilde{\gamma}_{12h} [n(\omega_{14},T_x) + 1] \rho_R
\label{eq:current_T1}
\end{multline}
and the corresponding heat flux into the system equals
\begin{equation}
q_{14;x} = (E_1-E_4)j_{14;x}.
\label{eq:heat_flux_T1}
\end{equation}
Probability/heat fluxes corresponding to the other transitions in the system of Fig.~\ref{fig:fridges}a can be calculated analogously.

Having identified the heat fluxes corresponding to the individual transitions, it is straightforward to calculate total heat fluxes from the individual reservoirs to the system by summing the heat fluxes caused by these reservoirs for the individual transitions. Using the formulas (\ref{eq:freqJ1})--(\ref{eq:freqJ5}) for the time derivatives $\dot{\rho}_{ii}$, it is straightforward to show that this identification of heat fluxes is in accord with their traditional derivation via the first law \cite{BulnesCuetara2016,Brandner2016}
\begin{equation}
\dot{U} = \frac{d}{dt}{\rm Tr}\left\{\mathcal{H}_S\tilde{\rho}\right\} = \sum_i E_i \dot{\rho}_{ii} = q_c + q_m + q_h.
\label{eq:first_law}
\end{equation}
Here $U$ denotes the average internal energy of the system and $q_x$ is the total heat entering the system from the reservoir at temperature $T_x$. 

We are interested in thermodynamic characterization  of the two refrigerator models depicted in Fig.~\ref{fig:fridges}. We assume that these refrigerators operate in a non-equilibrium steady state which is eventually reached by the system at long times after the transient relaxation period. In the steady state, the system density matrix is constant in time ($\dot{\rho}_{ij} = 0$), which implies the steady state version of the first law (\ref{eq:first_law}) $\dot{U} = q_c+q_m+q_h = 0$ together with the second law in terms of the total entropy produced per unit time, $\sigma$, solely determined by the entropy produced in the reservoirs as
\begin{equation}
\sigma = -\frac{q_c}{T_c} - \frac{q_m}{T_m} - \frac{q_h}{T_h} \ge 0.
\label{eq:tot_entropy}
\end{equation}
For the figures of merit of the refrigerators we take the ratios $q_c/\sigma$ of the cooling heat flux from the cold bath to the total entropy production and  $q_c/q_h$ of the cooling flux to the net energy flux feeding the system. The second law inequality (\ref{eq:tot_entropy}) together with the energy conservation impose an upper bound on $q_c/q_h$  --- using the condition $q_h > 0$ one finds 
\begin{equation}\label{eq:Eff1}
\frac{q_c}{q_h}  \le \frac{T_c}{T_h}\frac{T_h - T_m}{T_m - T_c}.
\end{equation}


\section{Refrigerators: Results}\label{Results}
Let us now investigate the above-defined quantities in the specific examples of the two refrigerators introduced in Fig.~\ref{fig:fridges}.
In the A-type fridge, each transition is caused by a single reservoir. According to Eq.~(\ref{eq:heat_flux_T1}), the energy/heat flux from the reservoir driving the specific transition is proportional to the product of energy difference and probability flux corresponding to this transition. The individual heat fluxes throughout the system are thus given by
\begin{eqnarray}
q_c &=& (E_4 - E_3)j
\label{eq:qcA}\\
q_m &=& - q_c - q_h 
\label{eq:qmA}\\
q_h &=& (E_1-E_4)j_{14} + (E_2-E_4)j_{24} = (E_1 - E_{4})j,
\label{eq:qhA}
\end{eqnarray}
where we have used Eq.~(\ref{eq:continuity}), stationary version of Eq.~(\ref{eq:first_law}), and the condition of the degeneracy of the levels $E_1=E_2$. 
All the heat fluxes throughout the system are thus proportional to the probability current $j$, which is the only quantity depending on the details of the system dynamics entering these fluxes. If $j>0$, the energy/heat drawn from the coldest bath is positive and the system works as a refrigerator.

The fact that all the heat currents are proportional to $j$ implies that for the A-type fridge both efficiencies (figures of merit) depend solely on the energies of the individual levels and on the reservoir temperatures: 
\begin{eqnarray}
\frac{q_c}{q_h} &=& \frac{E_4-E_3}{E_1 - E_4} \label{eq:qcqm}\\
\frac{q_c}{\sigma} &=& \frac{(E_4 - E_3) T_c T_h T_m}{E_3 T_h (T_c + T_m) - E_1 Tc (T_h + T_m) - E_4 (T_m - T_c) T_m }.\label{eq:qcsigma}
\end{eqnarray}
In particular, they do not depend on the values of the various rates $\gamma$. 

For the B-type refrigerator, the full probability current $j$ is for some of the transitions empowered by two different reservoirs. Using the notation of Eq.~(\ref{eq:heat_flux_T1}) and the associated argumentation, we find the following relations for the heat currents through the working medium of the B-type refrigerator:
\begin{eqnarray}
q_c &=& (E_4 - E_3)j_{43;c},
\label{eq:qcR}\\
q_m &=& - q_c - q_h,
\label{eq:qmR}\\
q_h &=& (E_1-E_4)j_{14;h} + (E_2-E_4)j_{24;h}.
\label{eq:qhR}
\end{eqnarray}
Now, the cooling flux $q_c$ is not directly proportional to the probability current $j$ and the efficiencies are complicated functions of the details of the system dynamics including the rates $\gamma$. 

In order to calculate the stationary probability current $j$, one must determine the steady state density matrix $\rho$ as the right eigenvector of the matrix $\mathcal{L}$ corresponding to the eigenvalue 0. While this can be found analytically for the both refrigerator models considered in this paper, the results are quite involved and we will not write them explicitly apart from certain limiting analytical results for the heat fluxes. In the general case we will just present numerical results graphically. 

\subsection{Cooling flux in the maximum coherence limit: approximate analytical study}
\label{sec:anal}

In this subsection, we will focus on the behavior of the A-type fridge in the simplified case of equal magnitudes of the elements of the dipole moment corresponding to the individual transitions to the degenerate doublet, $\gamma_{1h,m} = \gamma_{2h,m} = \gamma_{h,m}$. From the analysis in Sec.~\ref{Quantum} it follows that in this case the cooling fluxes obtained in the both types of fridges in the classical limit ($\gamma_{12h,m} = 0$) and in the maximum coherence limit ($\gamma_{12h,m} = \gamma_{h,m}$) coincide. To see this, it is enough to realize that in the classical case there exist two parallel paths from state $\left| 3\right>$ to state $\left| 4\right>$ with identical transition rates (see Fig.~\ref{fig:fridges}a, while in the maximum coherence case there is just one route from $\left| 3\right>$ to $\left| 4\right>$ with doubled rates as compared to the classical situation.

The limit of maximum coherence hence does not yield any gain in the cooling flux as compared to the classical case. The situation changes if one does not set precisely $\gamma_{12h,m} = \gamma_{h,m}$, but investigates the behavior of the cooling flux near this point. We find that this behavior is surprisingly complicated. The limit of the cooling flux $\gamma_{12h,m} \to \gamma_{h,m}$ depends on the direction from which the point $\gamma_{12h,m} = \gamma_{h,m}$ is approached in the $\gamma_{12h,m}$ plane (similar situation occurs in the studies \cite{Scully2011,Svidzinsky2011,Svidzinsky2012,Dorfman2013}).

To see this, consider the parameter regime where $n_h \gg n_c \gg n_m$ and $\gamma_c \to \infty$ (first take $n_h \to \infty$, then $\gamma_c \to \infty$, then $n_c \gg n_m$ and finally $n_c \gg 1$). If we first set $\gamma_{12m} = \gamma_m$ and then calculate the limit $\gamma_{12h} \to \gamma_h$ in this regime, we find that the resulting cooling flux is the same as in the classical limit:
\begin{equation}
q_c^0 = \frac{2 (E_4 - E_3) \gamma_m n_c}{1 + 4 n_c}.
\label{eq:qAclassic}
\end{equation}
On the other hand, if we first set $\gamma_{12h} = \gamma_h$ and then calculate the limit $\gamma_{12m} \to \gamma_m$, we find that the cooling flux is larger:
\begin{equation}
q_c = \frac{2 (E_4 - E_3) \gamma_m n_c}{1 + 3 n_c}.
\label{eq:qAcoh}
\end{equation}
For large $n_c$ the ratio $q_c/q_c^0$ equals to $4/3$ and thus the noise-induced coherence enhances the cooling flux of the A-type fridge by $\approx 33\%$.

Consider once again the system in the transformed basis, where the dynamics is described by a classical stochastic process  (see Fig.~\ref{fig:system_fridge}). As discussed above, one gains nothing by completely closing one of the transition channels from $\left| 3\right>$ to $\left| 4\right>$ and doubling transition rates for the other one. On the other hand, the formulas (\ref{eq:qAclassic}) and (\ref{eq:qAcoh}) demonstrate that an enhanced cooling flux can be obtained by closing just the transition channel between $\left| 1'\right>$ and $\left| 4\right>$  corresponding to the temperature $T_h$ (maximum coherence) and only almost closing the transition channel between $\left| 1'\right>$ and $\left| 3\right>$ corresponding to the temperature $T_m$ (nearly maximum coherence).

Similar behavior as described above occurs also for $\gamma_{1h,m} \neq \gamma_{2h,m}$ and for the B-type fridge as demonstrated numerically in the next subsection.

\subsection{Cooling flux in the maximum coherence limit: numerical study}
\label{sec:numerics}

In this subsection, we investigate whether the performance of both the A-type fridge and the B-type fridge can be enhanced by noise-induced coherence using reasonable values of physical parameters. In all the figures, we set $T_c = 34$ K, $T_m = 35$ K and $T_h = 5000$ K for the reservoir temperatures and $E_1 = E_2 = 0.01$ eV, $E_4 = 0.001$ eV and $E_3 = 0$ eV for the energies of the individual levels. We also always take $\gamma_c = 10^4 \gamma_0$. The parameter $\gamma_0 = (e\AA)^{2}\approx(5\, \mathrm{Debye})^{2}$ corresponds to a realistic value of the electric dipole element.

\begin{figure}
\centering
\begin{tikzpicture} 
\node (img1) {\includegraphics[width=0.90\columnwidth]{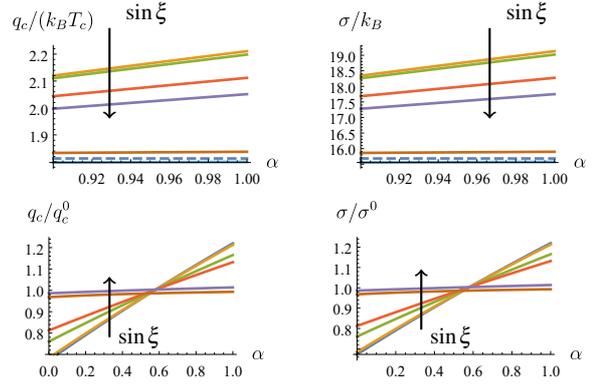}};
\draw [thick,<-] (2.5,1.0)--(2.5,2.2);
\node at (3.0,2.4) {$\sin\xi$};
\draw [thick,<-] (-2.5,1.0)--(-2.5,2.2);
\node at (-2.0,2.4) {$\sin\xi$};
\draw [thick,->] (-2.5,-1.9)--(-2.5,-1.1);
\node at (-2.1,-1.9) {$\sin\xi$};
\draw [thick,->] (1.6,-1.8)--(1.6,-1.0);
\node at (2.0,-1.9) {$\sin\xi$};
\end{tikzpicture}
	\caption{Performance of the A-type fridge in the parameter regime which allows for the description using a classical stochastic process. The individual full lines correspond to different values of the parameter $\xi$ from Eqs.~(\ref{eq:g12hfigs})--(\ref{eq:g12mfigs}) measuring the direction in which the maximum coherence limit is reached as $\alpha\to 1$. The arrows in the individual panels show the direction in which $\sin \xi$ gradually assumes the values 0, 0.001, 0.01, 0.02, 0.03 and 1. The dashed lines in the upper panels correspond to the classical case where $\gamma_{12h}=\gamma_{12m}=0$. Detailed description and the parameters used are given in Sec.~\ref{sec:numerics}.}
	\label{fig:qNonReal}
\end{figure}

Using these constants, we investigate the behavior of the A- and B-type refrigerators in two regimes. In the first one, we set $\gamma_{1h} = \gamma_{2h} = 5.7 \gamma_0$ and $\gamma_{1m} = \gamma_{2m} = 7.5 \gamma_0$ thus enabling to find a new basis in which the system dynamics is described by a classical stochastic process (see Sec.~\ref{Quantum}). In the second regime, we take $\gamma_{1h} = 5.7 \gamma_0, \gamma_{2h} = 4.0 \gamma_0, \gamma_{1m} = 7.5 \gamma_0$, and 
$\gamma_{2m} = 6.0 \gamma_0$ so that coherences and populations inevitably couple in the system dynamics. The time scales corresponding to the both sets of parameters are $\tau_{RWA} \approx 0.1 $ ps, $\tau_{R} \approx 0.2 $ ps, and $\tau_{S} \approx 3$ ms both for the A-type and for the B-type fridge. The assumptions of Sec.~\ref{sec:assumptions} used for the derivation of the master equation (\ref{eq:freq1})--(\ref{eq:freq6}) are thus fulfilled.

\begin{figure}
\centering
\begin{tikzpicture} 
\node (img1) {\includegraphics[width=0.90\columnwidth]{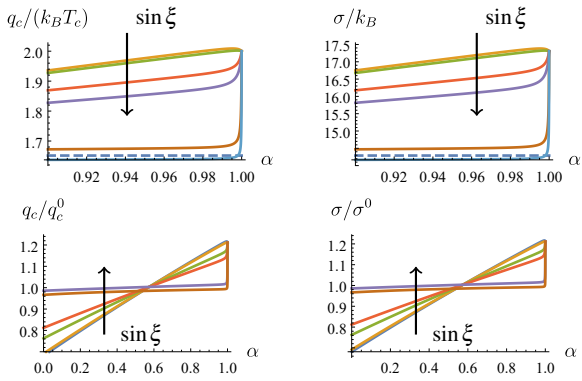}};
\draw [thick,<-] (2.4,1.0)--(2.4,2.1);
\node at (2.8,2.3) {$\sin\xi$};
\draw [thick,<-] (-2.2,1.0)--(-2.2,2.1);
\node at (-1.8,2.3) {$\sin\xi$};
\draw [thick,->] (-2.5,-1.9)--(-2.5,-1.0);
\node at (-2.0,-1.9) {$\sin\xi$};
\draw [thick,->] (1.6,-1.9)--(1.6,-1.0);
\node at (2.1,-1.9) {$\sin\xi$};
\end{tikzpicture} 
	\caption{Performance of the A-type fridge in the parameter regime which does not allow for the description using a classical stochastic process. The meaning of the individual lines and of the arrows is the same as in Fig.~\ref{fig:qNonReal}. Detailed description and the parameters used are given in Sec.~\ref{sec:numerics}.}
	\label{fig:qNonReal2}
\end{figure}

\begin{figure}
\centering
\begin{tikzpicture} 
\node (img1) {\includegraphics[width=0.90\columnwidth]{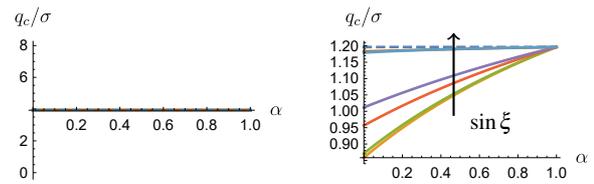}};
\draw [thick,->] (2.0,-0.3)--(2.0,0.8);
\node at (2.5,-0.4) {$\sin\xi$};
\end{tikzpicture} 
	\caption{Figure of merit $q_{c}/\sigma$ for the A-type fridge (left) and for the B-type fridge (right) in the parameter regime which allows for the description using a classical stochastic process. In the parameter regime which does not allow for the classical description, the figures of merit obtained for the two models of refrigerators behave almost identically. The meaning of the individual lines and of the arrow is the same as in Fig.~\ref{fig:qNonReal}. Detailed description and the parameters used are given in Sec.~\ref{sec:numerics}.}
	\label{fig:efficiencies}
\end{figure}

The behavior of the A-type fridge is depicted in Figs.~\ref{fig:qNonReal}, \ref{fig:qNonReal2} and left panel of Fig.~\ref{fig:efficiencies}. The performance of the B-type fridge is qualitatively almost the same as that of the A-type fridge and we thus show the corresponding numerical results only in the right panel of Fig.~\ref{fig:efficiencies}, where the performances of the two fridges differ.

The set of parameters which allows to describe the model dynamics by a classical stochastic process is used in Figs.~\ref{fig:qNonReal} and \ref{fig:efficiencies}. The parameter set which does not allow to describe the system dynamics by a classical stochastic process is used in Fig.~\ref{fig:qNonReal2}. In all the figures we plot dependencies of various quantities on two parameters $\alpha$ and $\xi$ which prescribe the coefficients $\gamma_{12h}$ and $\gamma_{12m}$ in the following way 
\begin{eqnarray}
\gamma_{12h} &=& [1 - (1-\alpha)\sin\xi]\sqrt{\gamma_{1h}\gamma_{2h}}\ ,
\label{eq:g12hfigs}\\
\gamma_{12m} &=& [1 - (1-\alpha\cos\xi]\sqrt{\gamma_{1m}\gamma_{2m}}\ .
\label{eq:g12mfigs}
\end{eqnarray}
The parameter $\sin\xi$ assumes the values 0, 0.001, 0.01, 0.02, 0.3 and 1 for the individual full lines from the top one to the lowest one (in the upper panels of Figs.~\ref{fig:qNonReal}, \ref{fig:qNonReal2}) or from the bottom one to the uppermost one (in the lower panels of Figs.~\ref{fig:qNonReal}, \ref{fig:qNonReal2} and in the right panel of Fig.~\ref{fig:efficiencies}). The dashed lines in all the figures correspond to the classical limit with $\gamma_{12h,m} = 0$.

In upper two panels of Figs.~\ref{fig:qNonReal} and \ref{fig:qNonReal2}, we show the cooling flux $q_c$ and the total entropy production $\sigma$ as functions of the parameter $\alpha$.  In the lower two panels, we show the ratios $q_c/q^0$ and $\sigma/\sigma^0$ of the individual full curves depicted in the upper panels to the corresponding dashed curve. The ratio $q_c/q^0$ depicts the gain in the cooling flux obtained due to the noise-induced coherence with respect to the classical situation. On the other hand, the ratio $\sigma/\sigma^0$ demonstrates  the fact that with increased cooling flux the noise-induced-coherence enhanced fridges produce larger amount of entropy than the classical ones. As a result, although the noise-induced coherence can enhance the cooling flux, it can not enhance any of the two efficiency measures $q_{c}/q_{h}$ and $q_{c}/\sigma$ as demonstrated in Fig.~\ref{fig:efficiencies} for the second one (the other measure $q_{c}/q_{h}$ behaves very similarly and we thus do not show it). 

Compared to the behavior of the A-type fridge depicted in Figs.~\ref{fig:qNonReal} and \ref{fig:qNonReal2}, the B-type fridge exhibits the following quantitative differences: i) The cooling heat flux $q_c$ is approximately two times smaller; ii) The entropy production $\sigma$ is approximately two times larger; iii) The ratio $q_c/q_c^0$ is slightly smaller;
iv) The ratio $\sigma/\sigma^0$ is slightly larger. The maximum gain in the cooling flux obtained by employing the noise-induced coherence, both in the A- and B-type fridges, is for the used parameters $\approx 25\%$ which is relatively close to the increase predicted in Sec.~\ref{sec:anal}. The biggest difference between the two models is found in Fig.~\ref{fig:efficiencies} --- while the efficiency of the A-type fridge does not depend on the varied parameters $\alpha$ and $\xi$ (as predicted in Eqs.~(\ref{eq:qcqm}) and (\ref{eq:qcsigma})), the efficiency of the B-type fridge changes with both. The most important finding of this figure is that, although the noise-induced coherence may enhance the cooling flux, the efficiency is always bounded by its classical value represented by the dashed line in the figure. This result is in accord with the findings of Refs. \cite{Brandner2016,Karimi2016,Brandner2017,Roulet2017}. These studies, however, treated models of periodically driven quantum machines, where coherences are not created by the noise, but rather by the driving itself. In such cases, the creation of coherences directly consumes part of the possible output power of the machine, thus decreasing its efficiency. Similar results to ours were also obtained for absorption refrigerators based on three linearly coupled harmonic oscillators \cite{Nimmrichter2017,Maslennikov2017}.

From the point of view of the two parameter sets used in the illustrations, we find that the curves for the parameters which allow to describe the model dynamics by a classical stochastic process in Fig.~\ref{fig:qNonReal} do not reach the same limiting value in the limit $\alpha \to 1$ for various $\xi$'s. This manifests the finding of Sec.~\ref{sec:anal} that the maximum-coherence limit $\gamma_{12h,m} \to \gamma_{h,m}$ of the cooling flux (and also of the entropy production)  depends on the direction from which the point $\gamma_{12h,m} = \gamma_{h,m}$ is approached in the $\gamma_{12h,m}$-plane. For the parameter set which does not allow for the classical description, this somewhat strange behavior vanishes. The curves corresponding to all values of the parameter $\xi$ attain the same limiting value for $\alpha \to 1$. While for the first set of parameters (``classical'') the maximum cooling flux is always obtained in the limit $\alpha \to 1$, in the second regime (``quantum'') the maximum cooling flux may be obtained for $\alpha < 1$ as shown in Fig.~\ref{fig:qNonReal2}, where the yellow curve peaks at $\alpha\approx 0.995$.


\section{Conclusions}\label{Conclusions}

We have studied in detail the performance of two model quantum absorption refrigerators with particular focus on the effects of noise-induced coherence. We have found  a completely analogous behavior to that previously reported for quantum heat engines, i.e.~coherences can enhance the power output but cannot increase the efficiency of engines and refrigerators. We have also put more attention to the distinction between ``kinematic'' coherences caused by the non-diagonal structure of the rate matrix, which can be under certain conditions fully removed by a suitable change of the basis, and truly dynamical quantum coherences revealed by an irreducible coupling of populations and coherences in the master equation. The noise-induced-coherence-based enhancement of the engine/refrigerator performance is invariably connected to the first type of the mechanism, while the influence of the true dynamical coherent coupling is far more subtle and in our view still requires further systematic studies before drawing a definite conclusion. 

We have tested the quantumness of the refrigerators behavior both indirectly by studying suitable quantities (figures of merit) characterizing their thermodynamic performance as well as directly by evaluating the Leggett-Garg inequalities with the same output. Namely, we have found that the true quantum case (featuring dynamical coherence) does not exhibit in the {\em stationary state} any significant qualitative difference from the case which can be reformulated as a classical stochastic process and both cases show even very similar (although surely not identical) quantitative behavior. The only qualitative difference between the two cases is the ambiguity of the stationary state in the maximum coherence limit for the ``classical'' case caused by the formation of a totally decoupled dark state, which disappears in the truly quantum case with dynamical coherence coupling. There is no doubt significant difference between the two cases for transient dynamics or for correlation functions but at the level of the stationary state quantities the differences are minor. This finding does by no means invalidate the existence of the noise-induced-coherence-enhancement mechanism, it just places its origin into the realm of static quantum phenomena which can be successfully emulated by classical processes with suitably chosen parameters as was already observed for quantum heat engines. True irreducible quantum dynamics does not appear to bring anything qualitatively new.         

\begin{acknowledgements}
We thank Clive Emary, Radim Filip, Karel Neto{\v c}n{\' y}, and Artem Ryabov for valuable discussions. This work was supported by the Czech Science Foundation (project No. 17-06716S). VH in addition gratefully acknowledges the support by the COST Action MP1209 and by the Alexander von Humboldt foundation.
\end{acknowledgements}




\begin{thebibliography}{10}
\providecommand{\url}[1]{{#1}}
\providecommand{\urlprefix}{URL }
\expandafter\ifx\csname urlstyle\endcsname\relax
  \providecommand{\doi}[1]{DOI \discretionary{}{}{}#1}\else
  \providecommand{\doi}{DOI \discretionary{}{}{}\begingroup
  \urlstyle{rm}\Url}\fi

\bibitem{Scully2003}
M.O. Scully, M.S. Zubairy, G.S. Agarwal, H.~Walther, Science
  \textbf{299}(5608), 862 (2003).
\newblock \doi{10.1126/science.1078955}.
\newblock \urlprefix\url{http://science.sciencemag.org/content/299/5608/862}

\bibitem{Dillenschneider2009}
R.~Dillenschneider, E.~Lutz, EPL (Europhysics Letters) \textbf{88}(5), 50003
  (2009).
\newblock \urlprefix\url{http://stacks.iop.org/0295-5075/88/i=5/a=50003}

\bibitem{Abah2014}
O.~Abah, E.~Lutz, EPL (Europhysics Letters) \textbf{106}(2), 20001 (2014).
\newblock \urlprefix\url{http://stacks.iop.org/0295-5075/106/i=2/a=20001}

\bibitem{Rosnagel2014}
J.~Ro\ss{}nagel, O.~Abah, F.~Schmidt-Kaler, K.~Singer, E.~Lutz, Phys. Rev.
  Lett. \textbf{112}, 030602 (2014).
\newblock \doi{10.1103/PhysRevLett.112.030602}.
\newblock
  \urlprefix\url{https://link.aps.org/doi/10.1103/PhysRevLett.112.030602}

\bibitem{Klaers2017}
J.~Klaers, S.~Faelt, A.~Imamoglu, E.~Togan, Phys. Rev. X \textbf{7}, 031044
  (2017).
\newblock \doi{10.1103/PhysRevX.7.031044}.
\newblock \urlprefix\url{https://link.aps.org/doi/10.1103/PhysRevX.7.031044}

\bibitem{Scully2011}
M.O. Scully, K.R. Chapin, K.E. Dorfman, M.B. Kim, A.~Svidzinsky, Proceedings of
  the National Academy of Sciences \textbf{108}(37), 15097 (2011).
\newblock \doi{10.1073/pnas.1110234108}.
\newblock \urlprefix\url{http://www.pnas.org/content/108/37/15097.abstract}

\bibitem{Svidzinsky2011}
A.A. Svidzinsky, K.E. Dorfman, M.O. Scully, Phys. Rev. A \textbf{84}, 053818
  (2011).
\newblock \doi{10.1103/PhysRevA.84.053818}.
\newblock \urlprefix\url{http://link.aps.org/doi/10.1103/PhysRevA.84.053818}

\bibitem{Svidzinsky2012}
A.~Svidzinsky, K.~Dorfman, M.~Scully, Coherent Optical Phenomena \textbf{1}, 7
  (2012).
\newblock \doi{doi:10.2478/coph-2012-0002}

\bibitem{Dorfman2013}
K.E. Dorfman, D.V. Voronine, S.~Mukamel, M.O. Scully, Proceedings of the
  National Academy of Sciences \textbf{110}(8), 2746 (2013).
\newblock \doi{10.1073/pnas.1212666110}.
\newblock \urlprefix\url{http://www.pnas.org/content/110/8/2746.abstract}

\bibitem{Tscherbul2014}
T.V. Tscherbul, P.~Brumer, Phys. Rev. Lett. \textbf{113}, 113601 (2014).
\newblock \doi{10.1103/PhysRevLett.113.113601}.
\newblock
  \urlprefix\url{https://link.aps.org/doi/10.1103/PhysRevLett.113.113601}

\bibitem{Killoran2015}
N.~Killoran, S.F. Huelga, M.B. Plenio, The Journal of Chemical Physics
  \textbf{143}(15), 155102 (2015).
\newblock \doi{10.1063/1.4932307}.
\newblock \urlprefix\url{https://doi.org/10.1063/1.4932307}

\bibitem{Xu2016}
D.~Xu, C.~Wang, Y.~Zhao, J.~Cao, New Journal of Physics \textbf{18}(2), 023003
  (2016).
\newblock \urlprefix\url{http://stacks.iop.org/1367-2630/18/i=2/a=023003}

\bibitem{Dodin2016}
A.~Dodin, T.V. Tscherbul, P.~Brumer, The Journal of Chemical Physics
  \textbf{145}(24), 244313 (2016).
\newblock \doi{10.1063/1.4972140}.
\newblock \urlprefix\url{https://doi.org/10.1063/1.4972140}

\bibitem{Su2016}
S.H. Su, C.P. Sun, S.W. Li, J.C. Chen, Phys. Rev. E \textbf{93}, 052103 (2016).
\newblock \doi{10.1103/PhysRevE.93.052103}.
\newblock \urlprefix\url{https://link.aps.org/doi/10.1103/PhysRevE.93.052103}

\bibitem{Streltsov2017}
A.~Streltsov, G.~Adesso, M.B. Plenio, Rev. Mod. Phys. \textbf{89}, 041003
  (2017).
\newblock \doi{10.1103/RevModPhys.89.041003}.
\newblock \urlprefix\url{https://link.aps.org/doi/10.1103/RevModPhys.89.041003}

\bibitem{Creatore2013}
C.~Creatore, M.A. Parker, S.~Emmott, A.W. Chin, Phys. Rev. Lett. \textbf{111},
  253601 (2013).
\newblock \doi{10.1103/PhysRevLett.111.253601}.
\newblock
  \urlprefix\url{https://link.aps.org/doi/10.1103/PhysRevLett.111.253601}

\bibitem{Chen2016}
H.B. Chen, P.Y. Chiu, Y.N. Chen, Phys. Rev. E \textbf{94}, 052101 (2016).
\newblock \doi{10.1103/PhysRevE.94.052101}.
\newblock \urlprefix\url{https://link.aps.org/doi/10.1103/PhysRevE.94.052101}

\bibitem{Levy2012}
A.~Levy, R.~Kosloff, Phys. Rev. Lett. \textbf{108}, 070604 (2012).
\newblock \doi{10.1103/PhysRevLett.108.070604}.
\newblock
  \urlprefix\url{https://link.aps.org/doi/10.1103/PhysRevLett.108.070604}

\bibitem{Correa2013}
L.A. Correa, J.P. Palao, G.~Adesso, D.~Alonso, Phys. Rev. E \textbf{87}, 042131
  (2013).
\newblock \doi{10.1103/PhysRevE.87.042131}.
\newblock \urlprefix\url{https://link.aps.org/doi/10.1103/PhysRevE.87.042131}

\bibitem{Brask2015}
J.B. Brask, N.~Brunner, Phys. Rev. E \textbf{92}, 062101 (2015).
\newblock \doi{10.1103/PhysRevE.92.062101}.
\newblock \urlprefix\url{https://link.aps.org/doi/10.1103/PhysRevE.92.062101}

\bibitem{Correa2014}
L.A. Correa, J.P. Palao, D.~Alonso, G.~Adesso, Scientific Reports \textbf{4},
  srep03949 (2014)

\bibitem{Silva2015}
R.~Silva, P.~Skrzypczyk, N.~Brunner, Phys. Rev. E \textbf{92}, 012136 (2015).
\newblock \doi{10.1103/PhysRevE.92.012136}.
\newblock \urlprefix\url{https://link.aps.org/doi/10.1103/PhysRevE.92.012136}

\bibitem{Gonzalez2017}
J.O. Gonz{\'a}lez, J.P. Palao, D.~Alonso, New Journal of Physics
  \textbf{19}(11), 113037 (2017).
\newblock \urlprefix\url{http://stacks.iop.org/1367-2630/19/i=11/a=113037}

\bibitem{Hofer2018}
P.P. Hofer, M.T. Mitchison, arXiv preprint arXiv:1803.06133  (2018)

\bibitem{Emary2014}
C.~Emary, N.~Lambert, F.~Nori, Reports on Progress in Physics \textbf{77}(1),
  016001 (2014).
\newblock \urlprefix\url{http://stacks.iop.org/0034-4885/77/i=1/a=016001}

\bibitem{Cohen-Tannoudji1977}
C.~Cohen-Tannoudji, B.~Diu, F.~Lalo{\"e}, \emph{Quantum mechanics}.
\newblock Quantum Mechanics (Wiley, 1977).
\newblock \urlprefix\url{https://books.google.cz/books?id=CnkfAQAAMAAJ}

\bibitem{Breuer2002}
H.P. Breuer, F.~Petruccione, \emph{The theory of open quantum systems} (Oxford
  University Press, 2002)

\bibitem{Lamb1947}
W.E. Lamb, R.C. Retherford, Phys. Rev. \textbf{72}, 241 (1947).
\newblock \doi{10.1103/PhysRev.72.241}.
\newblock \urlprefix\url{https://link.aps.org/doi/10.1103/PhysRev.72.241}

\bibitem{BulnesCuetara2016}
G.~Bulnes~Cuetara, M.~Esposito, G.~Schaller, Entropy \textbf{18}(12), 447
  (2016).
\newblock \doi{10.3390/e18120447}.
\newblock \urlprefix\url{http://www.mdpi.com/1099-4300/18/12/447}

\bibitem{Carmichael2009}
H.J. Carmichael, \emph{Statistical Methods in Quantum Optics 2: Non-Classical
  Fields} (Springer Science \& Business Media, 2009)

\bibitem{Joulain2016}
K.~Joulain, J.~Drevillon, Y.~Ezzahri, J.~Ordonez-Miranda, Phys. Rev. Lett.
  \textbf{116}, 200601 (2016).
\newblock \doi{10.1103/PhysRevLett.116.200601}.
\newblock
  \urlprefix\url{https://link.aps.org/doi/10.1103/PhysRevLett.116.200601}

\bibitem{Friedenberger2017}
A.~Friedenberger, E.~Lutz, EPL (Europhysics Letters) \textbf{120}(1), 10002
  (2017).
\newblock \urlprefix\url{http://stacks.iop.org/0295-5075/120/i=1/a=10002}

\bibitem{Brandner2016}
K.~Brandner, U.~Seifert, Phys. Rev. E \textbf{93}, 062134 (2016).
\newblock \doi{10.1103/PhysRevE.93.062134}.
\newblock \urlprefix\url{https://link.aps.org/doi/10.1103/PhysRevE.93.062134}

\bibitem{Karimi2016}
B.~Karimi, J.P. Pekola, Phys. Rev. B \textbf{94}, 184503 (2016).
\newblock \doi{10.1103/PhysRevB.94.184503}.
\newblock \urlprefix\url{https://link.aps.org/doi/10.1103/PhysRevB.94.184503}

\bibitem{Brandner2017}
K.~Brandner, M.~Bauer, U.~Seifert, Phys. Rev. Lett. \textbf{119}, 170602
  (2017).
\newblock \doi{10.1103/PhysRevLett.119.170602}.
\newblock
  \urlprefix\url{https://link.aps.org/doi/10.1103/PhysRevLett.119.170602}

\bibitem{Roulet2017}
A.~Roulet, S.~Nimmrichter, J.M. Arrazola, S.~Seah, V.~Scarani, Phys. Rev. E
  \textbf{95}, 062131 (2017).
\newblock \doi{10.1103/PhysRevE.95.062131}.
\newblock \urlprefix\url{https://link.aps.org/doi/10.1103/PhysRevE.95.062131}

\bibitem{Nimmrichter2017}
S.~Nimmrichter, J.~Dai, A.~Roulet, V.~Scarani, {Quantum} \textbf{1}, 37 (2017).
\newblock \doi{10.22331/q-2017-12-11-37}.
\newblock \urlprefix\url{https://doi.org/10.22331/q-2017-12-11-37}

\bibitem{Maslennikov2017}
G.~Maslennikov, S.~Ding, R.~Hablutzel, J.~Gan, A.~Roulet, S.~Nimmrichter,
  J.~Dai, V.~Scarani, D.~Matsukevich, arXiv preprint arXiv:1702.08672  (2017)

\bibitem{Lobejko2015}
M.~\L{}obejko, J.~\L{}uczka, J.~Dajka, Phys. Rev. A \textbf{91}, 042113 (2015).
\newblock \doi{10.1103/PhysRevA.91.042113}.
\newblock \urlprefix\url{https://link.aps.org/doi/10.1103/PhysRevA.91.042113}

\bibitem{Gardiner}
C.W. Gardiner, P.~Zoller, \emph{Quantum Noise}, 2nd edn. (Springer, 2000)

\end{thebibliography}



\appendix

\section{Leggett-Garg inequalities}
\label{Appx:LGI}

If the system dynamics cannot be described by a classical stochastic process, this is revealed by the breaking of the so called Leggett-Garg inequalities \cite{Emary2014}
for the two-time correlation function
\begin{equation}
C_{ij} = \left<Q(t_i)Q(t_j)\right>
\label{eq:two_time_CF}
\end{equation}
of a bounded observable $Q$ with possible outcomes lying in the interval $[-1,1]$, i.e. $|Q|\le 1$. Let 
\begin{equation}
K_n = C_{21} + C_{32} + \dots + C_{n(n-1)} - C_{n1}
\label{eq:LGS}
\end{equation}
denote the $n$-measurement Leggett-Garg string with $t_1 < t_2 < \dots < t_n$. Then the following inequalities for $n\geq 1$ must be fulfilled if the process in question is a classical stochastic one:
\begin{eqnarray}
- (2n+1) \le K_{2n+1} \le 2n-1, \label{eq:LGodd}\\
- 2n \le K_{2(n+1)} \le 2n.
\label{eq:LGeven}
\end{eqnarray}
The values of the Leggett-Garg strings in inequalities (\ref{eq:LGodd})--(\ref{eq:LGeven}) depend on the state of the system at the initial time $t_1$. By choosing a quantum superposition as the initial state, these transient Leggett-Garg inequalities can be easily broken for simplest systems. For example, considering a two-level system the maximum value of $K_3 = 1.5 > 1$ can be obtained  \cite{Lobejko2015}. 

If one chooses as the initial condition the stationary state, the inequalities (\ref{eq:LGodd})--(\ref{eq:LGeven}) reduce for equidistant measurement times to the condition
\begin{equation}
(n-1)\left<Q(t)Q(0)\right> - \left<Q((n-1)t)Q(0)\right> \le n-2\ .
\label{eq:LGstat}
\end{equation}
Thus the breaking of the stationary Leggett-Garg inequalities (\ref{eq:LGstat}) is proof for the fact that the system dynamics can not be described by a classical stochastic process even in the steady state into which the system eventually spontaneously relaxes regardless of the initial condition. 

\begin{figure}
\centerline{
    \begin{tikzpicture}[
      scale=0.5,
      level/.style={thick},
      virtual/.style={thick,densely dashed},
      trans/.style={thick,<->,shorten >=2pt,shorten <=2pt,>=stealth},
      classical/.style={thin,double,<->,shorten >=4pt,shorten <=4pt,>=stealth}
    ]  
		\draw[trans,red] (0cm,-3em) -- (-2.0cm,10.5em) node[midway,left] {\color{black} $\gamma$, $T_1$};
		\draw[trans,red] (0cm,-3em) -- (2.0cm,10.5em);
		\draw[trans,blue] (0.3cm,-3em) -- (2.3cm,10.5em) node[midway,right] {\color{black} $\gamma$, $T_1$, $T_2$};
    \draw[level] (1.5cm,-3em) node[right,above] {$\left|3\right>$} -- (-1.5cm,-3em) node[left,below] {}; 
		\draw[level] (-3.5cm,+10.5em) node[right,below] {} -- (-0.5cm,+10.5em) node[left,above] {$\left|1\right>$};
		\draw[level] (0.5cm,+10.5em) node[right,below] {} -- (3.5cm,+10.5em) node[left,above] {$\left|2\right>$};
    \end{tikzpicture}
}
\caption{V-type system used for testing quantumness of noise-induced coherence using the Leggett-Garg inequalities.}
\label{fig:3level_systemLG}
\end{figure}
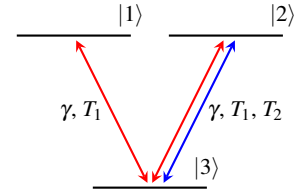

We have tested whether the dynamics of systems containing noise-induced coherence  can result in breaking the inequalities (\ref{eq:LGodd}) and (\ref{eq:LGstat}) using the specific system depicted in Fig.~\ref{fig:3level_systemLG}. In this system, the transitions from the lower level to both degenerate upper levels is caused by a reservoir at  temperature $T_1$. In addition to this, the transitions between the levels $\left| 2\right>$ and $\left| 3\right>$ can be also caused by another reservoir at temperature $T_2 \neq T_1$. We assume that the magnitudes of the dipole moment matrix elements corresponding to both transition channels equal to $\gamma$ and that the coefficient $\gamma_{12}$ which couples populations and coherences assumes its maximum value $\gamma_{12} = \gamma$. Using the procedure described around Eqs.~(\ref{eq:3level_ini})--(\ref{eq:3level_end}), we find that this system is described by the set of dynamical equations
\begin{eqnarray}
\dot{\rho}_{11} &=&  n_1 \rho_{33} - (n_1+1)\rho_{11} -  (n_1+1) \rho_R ,
\label{eq:3level_ini2}\\
\dot{\rho}_{22} &=& (n_1+n_2)\rho_{33} - (n_1+n_2+2)\rho_{22} -  (n_1+1) \rho_R,\\
\dot{\rho}_{33} &=& (n_1+1)\rho_{11} +  (n_1+n_2+2)\rho_{22} - (2n_1+n_2)\rho_{33} 
\nonumber
\\
&+&  2(n_1+1)\rho_R,
 \\
\dot{\rho}_R &=& n_1\rho_{33} - \frac{n_1+1}{2}(\rho_{11} + \rho_{22})- \left[2(n_1+1) + \frac{1}{2}(n_2+1)\right]\rho_R,
\label{eq:3level_end2}
\end{eqnarray}
with $\dot{\rho}\equiv d\rho(t)/\tilde{\gamma} dt$ and the  same meaning of the coefficients $n_{1,2}$ and $\tilde{\gamma}$ as in Eqs.~(\ref{eq:freq1})--(\ref{eq:freq6}). Note that the bath at $T_2$ which couples only with one transition channel does not increase the coupling between populations and coherences, but it causes faster decay of coherences. At long times, the system described by Eqs.~(\ref{eq:3level_ini2})--(\ref{eq:3level_end2}) reaches a non-equilibrium steady state with non-zero noise-induced coherence. As can be checked by substituting parameters of the present model ($\gamma_{1h}=\gamma_{2h}=\gamma_{12h}=\gamma_{2m}=\gamma$, $\gamma_{1m}=\gamma_{12h}=0$) to Eqs.~(\ref{eq:transQ1})--(\ref{eq:transQ3}), the coupling to the two reservoirs is chosen in such a way that there exists no basis in the degenerate subspace $\{|1\rangle,|2\rangle\}$ which would lead to decoupling of coherences and populations in the master equation.

Using the quantum regression theorem \cite{Gardiner}, we have calculated both the Leggett-Garg string $K_3$ and the stationary time-correlation function $2\left<Q(t)Q(0)\right> - \left<Q(2t)Q(0)\right>$ for the set of observables of the type $Q_1 = (\cos \theta \left| 1\right> + \sin{\theta} \left| 2\right>)(\cos \theta \left< 1\right| + \sin{\theta} \left< 2\right|)$ and $Q_2 = |1\rangle\langle 1|+|2\rangle\langle 2|+|3\rangle\langle 3| - Q_1$ exploring a large part of the model parameter space. While the transient Leggett-Garg inequality (\ref{eq:LGodd}) can be indeed broken by the present system if one choses a suitable initial condition, we were not able to break the stationary Leggett-Garg inequality (\ref{eq:LGstat}). This suggests, that the effect of noise-induced coherence in the steady state can be mimicked by a classical stochastic dynamics. This is actually in accord with the results of the study \cite{Creatore2013} where performance of a heat engine similar to that found in noise-induced-coherence works \cite{Scully2011}, \cite{Svidzinsky2011}, \cite{Svidzinsky2012}, \cite{Dorfman2013} has been achieved using diagonal elements of the density matrix only. For an example of a quantum engine which breaks the Legget-Garg inequalities, we refer to Ref.~\cite{Friedenberger2017}.

\end{document}